\documentclass [preprint,aps]{revtex4}
\usepackage{epsfig}
\usepackage{amssymb}

\begin{document}
\title{
EXPERIMENTS WITH PARAMETRIC X-RAY RADIATION (PXR) FROM
NON-RELATIVISTIC ELECTRONS}

\author{V.G.Baryshevsky, K.G.Batrakov, I.D.Feranchuk, A.A.Gurinovich,
A.O.Grubich, A.S.Lobko, A.A.Rouba, B.A.Tarnopolsky, P.F.Safronov,
V.I.Stolyarsky, A.P.Ulyanenkov}

\address{
Institute for Nuclear Problems, Belarus State University \\ 11
Bobrujskaya Str., Minsk 220050, Belarus }

\begin{abstract}

Interaction of non-relativistic electrons with single crystal target may
produce coherent x-rays. That is the result of interference between two
known x-ray generation mechanisms having orientational behavior, namely
parametric x-rays and coherent {\it Bremsstrahlung}. Experiments aimed to PXR research were
performed with 50-100 keV electrons and its distinctive features were
observed. Requirements to the experimental set-up, detector instrumental
response, and targets as well as experiment geometry are discussed in
detail. Series of PXR spectra in various conditions were recorded and their
distinctive features were observed. Tuning of radiation frequency with
crystal-target rotation was observed for the first time for low energy
electrons. Dependence of the x-ray frequency on the beam energy was
detected. Soft PXR peak with energy below 1~keV was observed for the first
time. Possible applications of PXR for structure analysis and
crystallography are discussed. These results are obtained in the framework
of ISTC project {\#}B626.

\end{abstract}
\maketitle

\section{INTRODUCTION}

Parametric x-rays produced by non-relativistic electrons passing
through a single crystal target were described in \cite{1} on the
basis of general theory of PXR \cite{2}. Due to the fact that the
angular distributions of x-rays originated from all radiation
mechanisms are almost isotropic in this case, peaks produced by
coherent effects, i.e. parametric x-rays and coherent {\it
Bremsstrahlung} (CBS), are the result of PXR and CBS interference
and may be observed on the intensive uniform background. Their
shape and intensity depend on the target thickness and the
detector spectral resolution. Some experiments with
non-relativistic electrons are currently known \cite{3}-\cite{5},
but their authors outlined the effect taking into consideration
only CBS mechanism. Model \cite{1} presumably describes x-rays
from non-relativistic electrons in crystal targets more
adequately.

To study x-ray properties followed from the theory \cite{1}, we
have performed experiments focused on the PXR research with
50$\div$100~keV electrons of the electron microscope. Let us
briefly consider requirements of the PXR experimental observation.
In general case a charged particle passing though matter undergoes
elastic and inelastic collisions with the atoms of a medium. As a
result, the angular and energy distributions of electrons change
during beam passing through a medium. Influence of electron
scattering on the spectral-angular distribution becomes
significant if either the width of velocity distribution
$\frac{\Delta \nu }{\nu }$ or the width of angular distribution
$\Delta \theta _\ast $ becomes larger or equal to the width of the
spectral line: $\frac{\Delta \nu }{\nu }\ge \frac{1}{kL_\ast }$ or
$\Delta \theta _\ast \ge \frac{1}{kL_\ast }$, where
$k=\frac{\omega }{c}$ is the wave-vector of the radiated photon,
$L_\ast $ is the electron path in the crystal.

Angles of multiple scattering and influence of inelastic
scattering have been estimated for several crystals at low
electron energy. It was obtained, that to observe coherent
x-radiation in wide frequency range the rigid requirements for
target thickness and electron beam parameters should be fulfilled.
They are: the target thickness should not exceed $\sim $0.5~$\mu
$m and the energy of electron beam should be above 50-60~keV (in
this energy range angle of multiple scattering $<\theta_{s}^2>
\sim E^{-2})$. Initial energy and angular dispersion of the
electron beam should be less than 10$^{-1}$.

We propose to detect x-ray photons, which are radiated at small
sliding angle $\psi \ll 1$ with respect to the target surface in
the direction, which is opposite to electron velocity. In this
case the detector would register the photons from the electron
pass $\sim \psi/(k \chi^{\prime \prime})$ (if $\psi/(k
\chi^{\prime \prime}) < 1$), here $1/(k \chi^{\prime \prime})$ is
approximately equal to the absorption length of radiation in a
crystal, $\chi^{\prime \prime}$ is the imaginary part of crystal
susceptibility. Therefore, a part of electron beams with less
dispersion ($<\theta _{s}^2>\sim L_\ast$) radiates. This
experiment can be realized for such target and radiation
frequency, which correspond to small absorption length. For
example, absorption length for LiH crystal is $\sim $1~$\mu $m for
10 $\AA$ radiation wavelength. Then, at the angle $\psi \sim $0.1
($\sim 5.73^\circ$) one can observe radiation from the 0.1~$\mu $m
pass.

The PXR photons can be radiated at all directions. The frequency of PXR
photons depends not only on the angle between the electron velocity and
crystallographic plane {\it $\theta $}$_{H}$ but also on the angle $\theta $ between the
electron velocity and direction of registration
\begin{eqnarray}
\omega_{H}^{(n)}(\theta)=\frac{2 \pi v \cos{\theta_H}}{d (1-v/c
\cos{\theta})} n,~ n=1,2,... \label{1}
\end{eqnarray}
where $d$ is the interplanar interval, $v$ is the electron
velocity, $c$ is the speed of light. When the observation angle
$\theta $ is fixed, the PXR spectral width is defined by the
length of the coherent interaction of the electron in crystal,
that is before the multiple scattering becomes essential. So, the
effective observation of PXR with the non-relativistic electrons
is possible for the single crystal films with the thickness lower
than effective scattering length, $L<L_{sc}$. In this case the
relative spectral width of the PXR "harmonic" is defined by the
coherent interaction of the electrons with crystal and can be
calculated as follows:

\begin{eqnarray}
\frac{\Delta \omega_0}{\omega}=\frac{v}{L
\omega_{H}^{(n)}(\theta)}n
 \label{2}
\end{eqnarray}

\section{EXPERIMENT}

\begin{figure}
\centering
\includegraphics [scale=1] {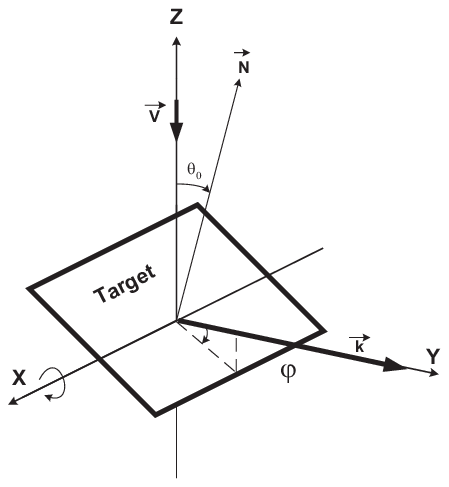}
\includegraphics [scale=0.6] {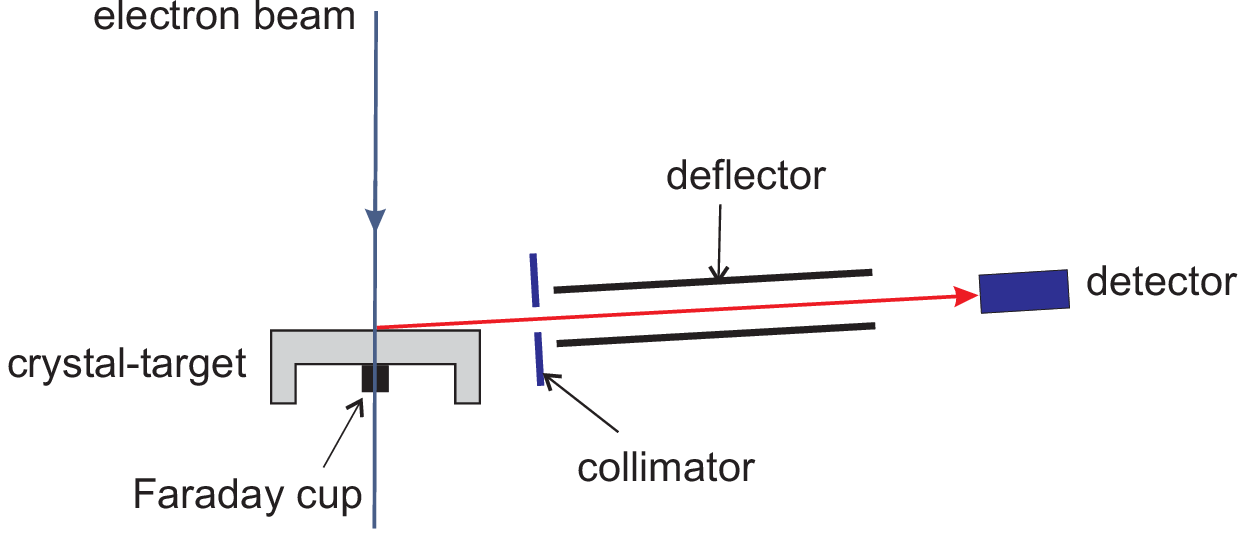}
\caption {Scheme and layout of the experiment} \label{Fig1}
\end{figure}

Taking into account conditions described above, we have performed
the experiment outlined in Fig. {\ref{Fig1}}. Narrow electron beam
falls on the surface of the single crystal target, $\vec{v}$ is
the velocity of electron beam ($\vec{v}\uparrow \downarrow OZ$).
Axis $OX$ is coplanar the target surface, $\vec{N}$ is orthogonal
to the crystal surface. Angle $\Theta_{0}$ is the angle between
$\vec{N}$ and $OZ$, which determines crystal rotation around the
axis $OX$. At $\Theta_{0}=0$ the target surface is orthogonal to
the velocity of electron beam. At crystal rotation around the axis
$OX$, $\vec{N}$ moves in the plane $ZY$. Detector window is placed
at the angle $\sim \pi /2$ to the velocity of electrons in the
direction $OY$.

To provide requirements to the beam, we have applied electron beam of an
electron microscope. Electron beam injector was thoroughly tested
for stability and reliability of beam shaping systems and to optimize beam
parameters for CXR observation. Some measured beam parameters are as
follows:

\begin{itemize}
\item electron energy 50-100~keV
\item relative instability of accelerating voltage 2*10$^{-5}$
\item brightness 7*10$^{4}$ À/m$^{2}$ sr.
\end{itemize}
Careful selection of optimal currents of lenses and voltage between Venelt
cylinder and anode minimized parasitic scattering inside the microscope and
provide minimal x-ray background inside the experimental area. To reduce
background caused by electron scattered at the target, we have applied
electrostatic deflector made of two brass plates of the 50~mm length
positioned in the input window of the detector. The six kilovolts voltage
was applied between to them that decrease background rate significantly.

We have used Si(Li) detector with thermoelectric cooling and Si(Li) detector
with cryostat both with 20~mm$^{2}$ sensitive areas and thin polymer windows
for soft x-ray detection supplied by BSI (Riga, Latvia). Energy resolution
of detectors was evaluated as $\sim $170~eV. Detector angular aperture was
0.2~mrad. ORTEC 2056-C 4096-channel analyzer collected detector output data.
Off-line spectra processing was performed after spectra transferring to the
computer. It involved calibration, smoothing, baseline subtraction, and
fitting by the set of Gauss peaks.

\begin{figure}
\centering
\includegraphics [scale=0.6] {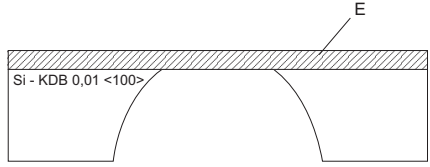}
\caption {Sketch of Si (100) target of sub-micron thickness, E =
epitaxial layer} \label{Fig2}
\end{figure}

Analysis above shows that optimal target thickness should be below
half of micron. Applied target (Fig.{\ref{Fig2}}) is the silicon
crystal substrate of 2x2~mm dimensions and $\sim $200~$\mu $m
thickness with $\sim $0.5~$\mu $m thickness membrane of 1.0~mm
diameter. Basic plane has (100) or (111) orientation. Membrane
material is layer of pure epitaxial Si of $\sim $0.9--1.0~$\mu $m
thickness deposited on substrate of heavily doped p$^{+}$ Si of
KDB 0.01~$<$100$>$ grade. Choice of such structure was determined
by electrochemical etching technique, in which pure epitaxial Si
serves as termination layer. For membrane of other thickness one
should take structures with epitaxial layer thickness close to
desired one. Precise membrane thickness adjustment can be
performed by ion-beam etching with $\sim $10-15~nm/min rate.

We have used relatively simple technique to measure thickness of such
ultra-thin Si targets. As membranes with thickness of about micron and below
are going to be semi-transparent in visible light range, one can record
their optical transmittance spectra. At this spectra interference fringes
can be observed and than thickness may be calculated. Dispersion of measured
thickness values normally is below 5{\%}.

\section{PRELIMINARY RESULTS AND DISCUSSION}

Experiments described below were performed with (111) Si target of $\sim
$410~nm thickness. Working current of $\sim $150~nA was chosen to provide
count rate at detector below 3~kHz and avoid peak distortion. Data
acquisition time was 5,000~seconds for the majority of spectra. Some spectra
were recorded at 10,000~seconds, but due to equipment instabilities obtained
spectra did not have considerable improvement.

\begin{figure}
\centering
\includegraphics [scale=1] {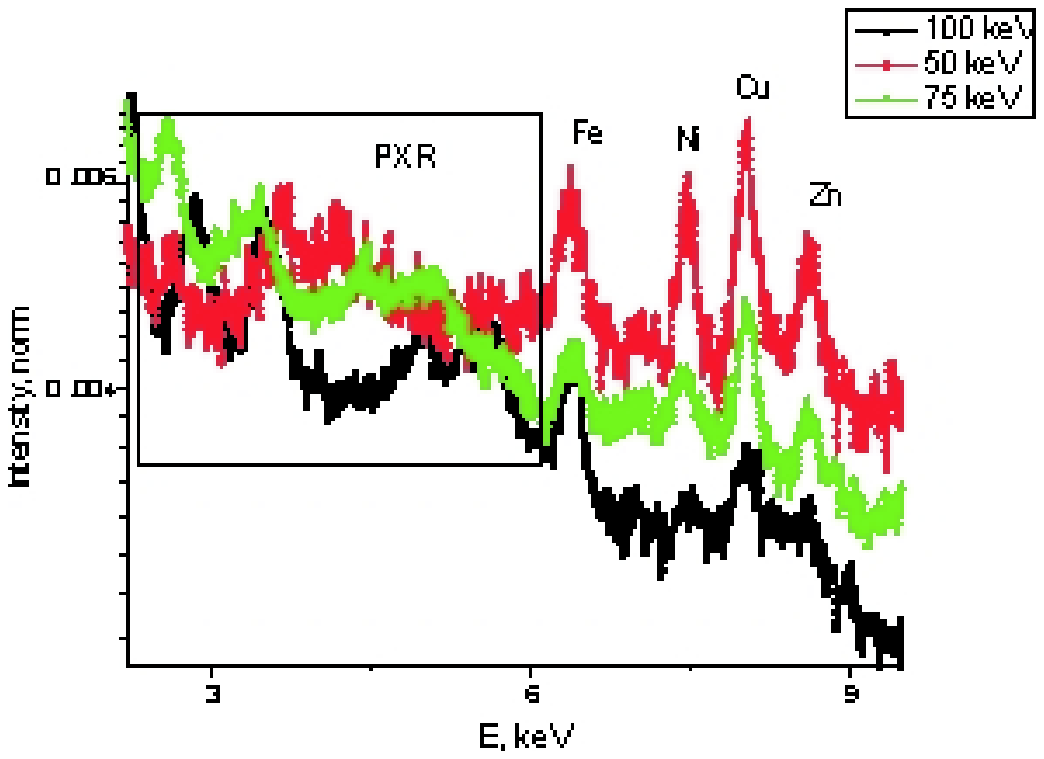}
\caption {Raw PXR spectra at 50, 75, and 100~keV electron energy}
\label{Fig3}
\end{figure}

Fragments of raw (just smoothed by Savitsky-Golay algorithm)
normalized PXR spectra are shown in Fig.{\ref{Fig3}}. One can see
dependence of PXR peaks frequency on beam energy. PXR frequency is
increasing with beam energy increase in conformity with formula
(\ref{1}). Group of characteristic x-ray peaks in right part of
spectra correspond to the microscope constructional materials
excited by electrons scattered by the target. Characteristic peak
of Si locates at $\sim $1.73~keV and it is not shown in picture
because it is very intensive in comparison with other peaks.

\begin{figure}
\centering
\includegraphics [scale=1] {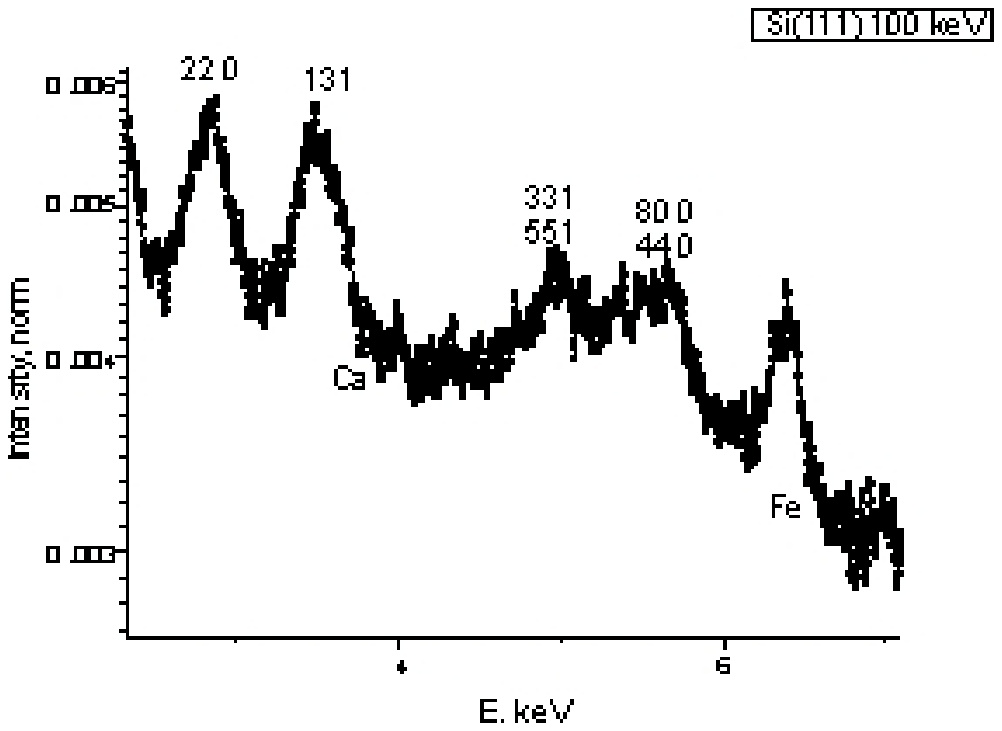}
\caption {Raw PXR spectra at 50, 75, and 100~keV electron energy}
\label{Fig4}
\end{figure}

After more detailed PXR spectra examinations we can attribute
peaks to corresponding crystallographic reflexes,
Fig.{\ref{Fig4}}. Most intensive peak in 2.5-2.9~keV region is
related to (220) reflexes with various indexes permutations.
Reflex (131) can correspond to peak at 3.5~keV. Region 4.9-5.2~keV
can be attributed to (331) and (511) including permutations.
Finally, region 5.4-6.0~keV is corresponding to (800) and (440)
reflexes.

\begin{figure}
\centering
\includegraphics [scale=1] {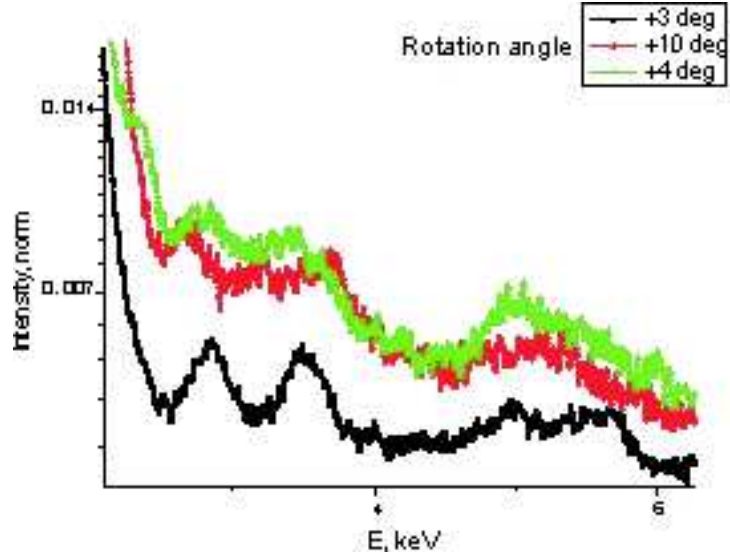}
\caption {PXR frequency tuning with crystal target rotation}
\label{Fig5}
\end{figure}

As followed from Eq.(\ref{1}), PXR frequency must depend on beam
incidence angle. Experimentally observed frequency sifts are shown
in Fig.{\ref{Fig5}}. Basically, peaks should be splitted and
distance must increase depending on incidence angle. At 3~degrees
incidence angle split should be about 0.2~keV, at 10~degrees it
may be as high as 0.7~keV. One can see this split in some spectra.
Changes in spectra connected with incidence angle change are close
to calculated values.

\section{CONCLUSION}

Series of PXR spectra in various conditions were recorded and
their distinctive features were observed. Tuning of radiation
frequency with crystal-target rotation was observed for the first
time for low energy electrons. Dependence of the x-ray frequency
on the beam energy was detected. Despite of their relatively low
quantum yield, they can be considered as prospective source for
structure analysis and crystallography.

We have obtained and used big amount of targets made of various materials.
Spectra of acceptable quality were measured with only a few thin Si
membranes. Expecting application of PXR of non-relativistic electrons as
base for a source of tunable x-rays, problem of thin single crystal membrane
target must be a matter of high-tech challenge.

These results are obtained in the framework of ISTC project
{\#}B626.

\end{document}